\documentclass[tog]{acmsiggraph}

\usepackage{subfigure,overpic}
\usepackage{ulem,soul} 
\usepackage{enumerate}
\usepackage{amssymb,mathtools}
\usepackage{amsmath}
\usepackage{multirow}
\usepackage{algorithm,algorithmic}
\usepackage{xcolor}
\usepackage{enumitem}
\setenumerate[1]{itemsep=0pt,partopsep=0pt,parsep=\parskip,topsep=1pt}
\setitemize[1]{itemsep=0pt,partopsep=0pt,parsep=\parskip,topsep=2pt}
\setdescription{itemsep=0pt,partopsep=0pt,parsep=\parskip,topsep=5pt}

\newcommand{\sArt}{state-of-the-art~}
\newcommand{\Rows}[1]{\multirow{2}{*}{#1}}
\newcommand{\figref}[1]{Fig.~\ref{#1}}
\newcommand{\tabref}[1]{Tab.~\ref{#1}}
\newcommand{\equref}[1]{(\ref{#1})}
\newcommand{\secref}[1]{Sec.~\ref{#1}}
\newcommand{\algref}[1]{Alg.~\ref{#1}}

\newcommand{\myPara}[1]{\vspace{-.08in}\paragraph{#1.}}

\newcommand{\CRF}{{\sc crf}}
\newcommand{\NYU}{{\sc nyu}}
\newcommand{\CORE}{{\sc core}}

\newcommand{\etal}{{\it et al.}}
\newcommand{\GMM}{{\sc gmm}}
\newcommand{\cf}{{c.f.~}}

\newcommand{\tbl}{\vspace{-.08in}\caption}
\renewcommand{\tabcolsep}{1.1mm}

\TOGonlineid{45678}
\TOGvolume{0}
\TOGnumber{0}
\TOGarticleDOI{1111111.2222222}
\TOGprojectURL{http://mmcheng.net/imagespirit/}
\TOGvideoURL{http://mmcheng.net/mftp/Papers/ImageSpirit.mp4}
\TOGdataURL{http://mmcheng.net/imagespirit/}
\TOGcodeURL{http://mmcheng.net/imagespirit/}

\graphicspath{{Imgs/}}

\title{ImageSpirit: Verbal Guided Image Parsing}

\author{Ming-Ming Cheng$^{1}$ \quad Shuai Zheng$^{1}$\thanks{Joint first author. Project page: http://mmcheng.net/imagespirit/.}
    \quad Wen-Yan Lin$^{2}$ \quad  Vibhav Vineet$^{2}$ \quad Paul Sturgess$^{2}$ \quad Nigel Crook$^{2}$ \\
	Niloy J. Mitra$^{3}$ \quad  Philip Torr$^{1}$ \\
    $^1$University of Oxford \quad $^2$Oxford Brookes University \quad $^3$University College London
}
\pdfauthor{Cheng et al.}


\begin{document}

\teaser{
  \vspace{-.15in}
  \subfigure[Inputs: an image and object/attributes potentials \protect \cite{shotton2009textonboost} ]{\includegraphics[height=.221\textwidth]{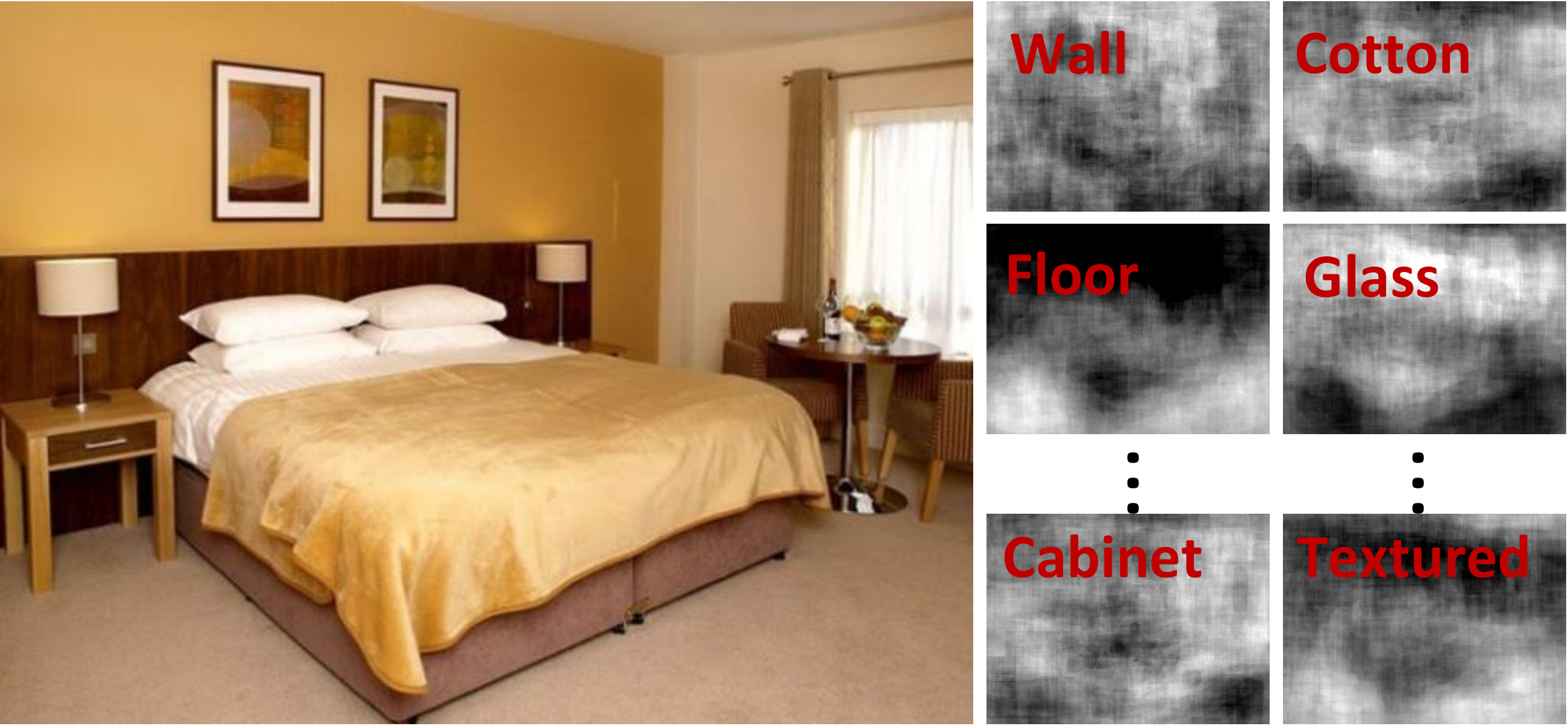}} \hfill
  \subfigure[Automatic scene parsing results]{\includegraphics[height=.221\textwidth]{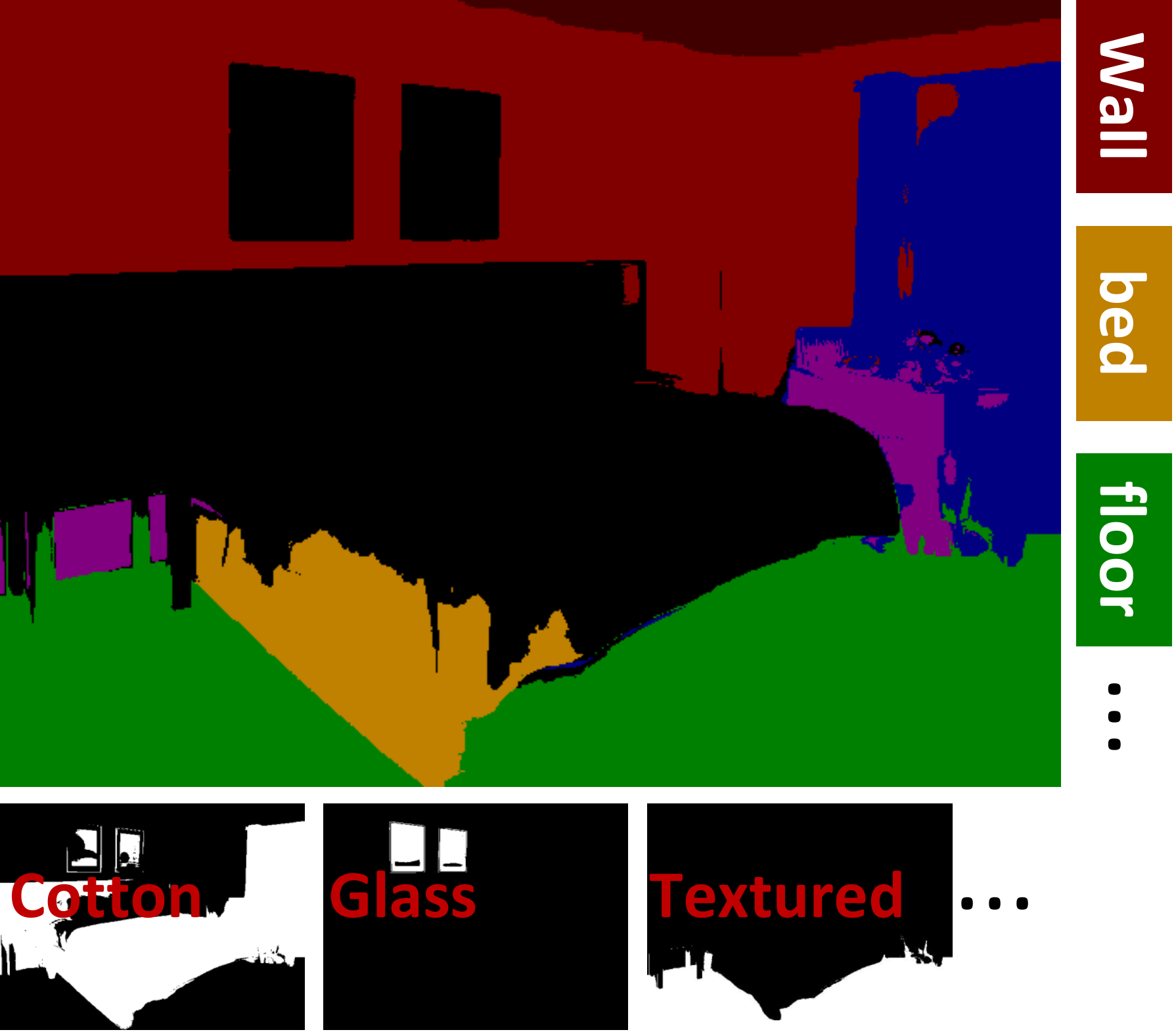}} \hfill
  \subfigure[Natural language guided parsing]{\includegraphics[height=.221\textwidth]{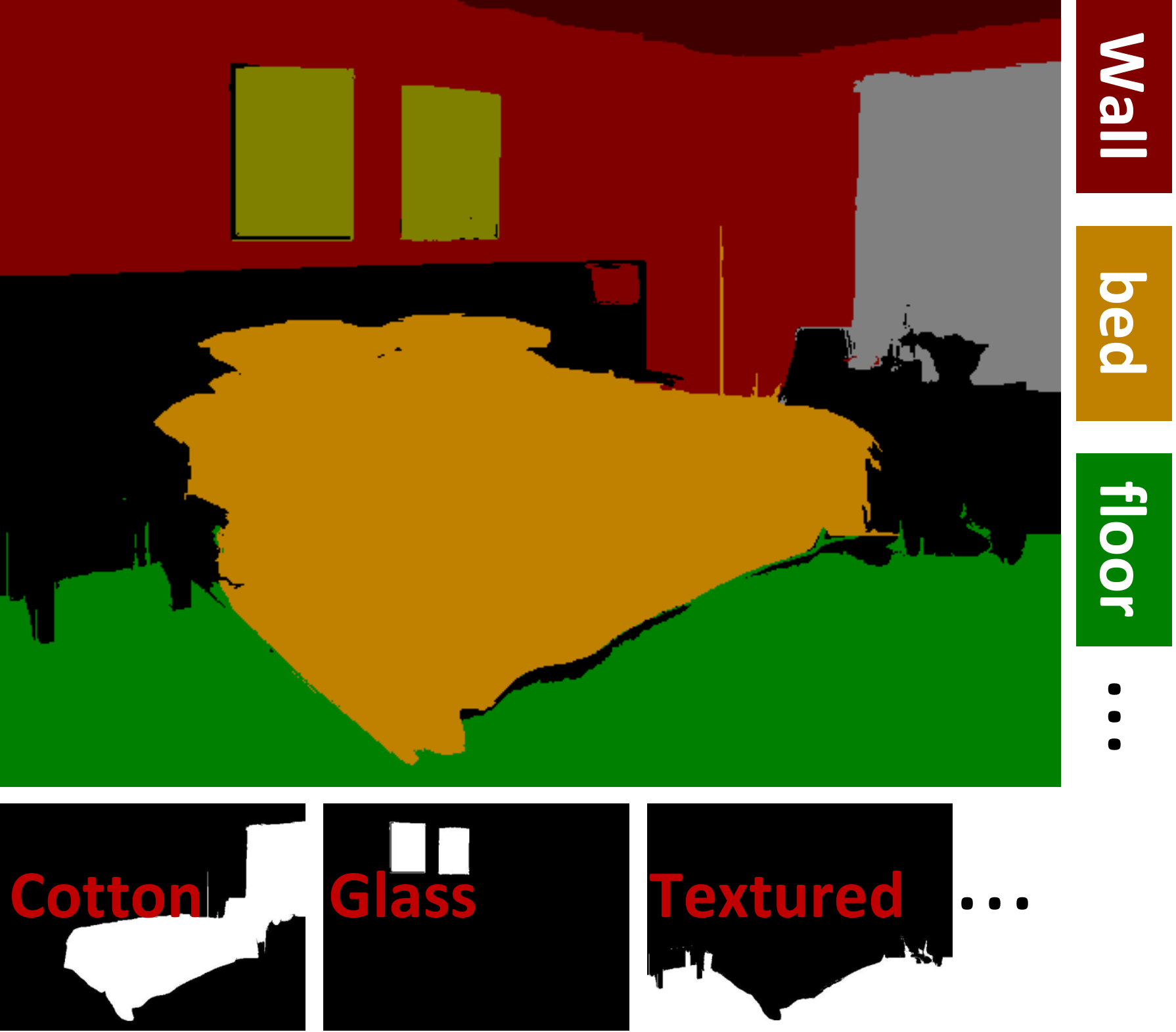}}
  \\ \vspace{-.18in}
  \caption{(a) Given a source image downloaded from the Internet,
    our system generates multiple weak object/attributes cues.
    (b) Using a novel multi-label \CRF, we generate an initial per-pixel object and attribute labeling.
    (c) The user provides the verbal guidance:
    `Refine the cotton bed in center-middle',
    `Refine the white bed in center-middle',
    `Refine the glass picture',
    `Correct the wooden white cabinet in top-right to window'
    allows re-weighting of \CRF~terms to generate,
    at interactive rates, high quality scene parsing result.
  } \label{fig:teaser}
}

\maketitle

\begin{abstract}

Humans describe images in terms of nouns and adjectives while algorithms operate on images represented as sets of pixels.
Bridging this gap between how humans would like to access images versus their typical representation
is the goal of image parsing, which involves assigning object and attribute labels to pixel.
In this paper we propose treating nouns as object labels and adjectives as visual attribute labels.
This allows us to formulate the image parsing problem as one of jointly estimating per-pixel object
and attribute labels from a set of training images.
We propose an efficient (interactive time) solution. Using the extracted labels as handles,
our system empowers a user to verbally refine the results.
This enables hands-free parsing of an image into pixel-wise object/attribute labels that correspond to human semantics.
Verbally selecting objects of interests enables a novel and natural interaction modality that
can possibly be used to interact with new generation devices (e.g. smart phones, Google Glass, living room devices).
We demonstrate our system on a large number of real-world images with varying complexity.
To help understand the tradeoffs compared to traditional mouse based interactions,
results are reported for both a large scale quantitative evaluation and a user study.

\end{abstract}


\keywordlist


\TOGlinkslist



\section{Introduction}
Humans describe images in terms of language components such as nouns
(e.g. bed, cupboard, desk) and adjectives (e.g. textured, wooden).
In contrast, pixels form a natural representation for computers~\cite{VisualAttributesNIPS2007}.
Bridging this gap between our mental models and machine representation
is the goal of image parsing~\cite{Imageparsing/IJCV2005,Tighe/ijcv2013}.
The goals of this paper are two-fold: develop a new automatic image parsing model that can handle attributes (adjectives) and objects (nouns), explore how to interact verbally with this parse in order to improve the results.
This is a difficult problem.
Whilst to date, there exists a large number of automated image parsing techniques
\cite{AssociativeHCRFICCV2009,shotton2009textonboost,krahenbuhl2011efficient,kulkarni2011baby,understandingscenesICCV2011},
their parsing results often require additional refinement before being useful
for applications such as image editing.
In this paper, we propose an efficient approach that allows  users
to produce high quality image parsing results from verbal commands.
Such a scheme enables hands-free parsing of an image into pixel-wise object and attribute labels
that are meaningful to both humans and computers.
The speech (or speech \& touch) input is useful for
the new generation of devices such as smart phones, Google Glass, consoles and living room devices,
which do not readily accommodate mouse interaction.
Such an interaction modality not only enriches how
we interact with the images,
but also provides an important interaction capability for applications
where non-touch manipulation is crucial \cite{hospital}
or hands are busy in other ways~\cite{arMantenance}.

We face three technical challenges in developing verbal guided\footnote{
We use the term verbal as a short hand to indicate word-based, i.e.,
nouns, adjectives, and verbs. We make this distinction as we focus
on semantic image parsing rather than speech recognition or natural language processing.}
image parsing:
(i)~words are concepts that are difficult to
translate into pixel-level meaning;
(ii)~how to best update the parse using verbal cues;
and (iii)~ensuring the  system responds at interactive rates.
To address the first problem, we treat nouns as objects and adjectives as attributes.
Using training data, we obtain a score at each pixel for each object and attribute,
e.g. \figref{fig:teaser}(a).
These scores are integrated through a novel, multi-label factorial conditional random field~(\CRF)
model\footnote{We substantially extended this model in \cite{AttObj2014} to include
hierarchical relations between regions and pixels, improved attribute-object relationship learning, etc.}
that jointly estimates both object and attribute predictions.
%
%
We show how to perform inference on this model to obtain an initial scene parse as demonstrated in \figref{fig:teaser}(b).
This joint image parsing with both objects and attributes provides verbal handles to
the underlying image which we can now use for further manipulation of the image.
Furthermore, our modeling of the symbiotic relation between attributes and objects
results in a higher quality parsing than considering each separately~\cite{AssociativeHCRFICCV2009,krahenbuhl2011efficient}.
To address the second problem,
we show how the user commands can be used to update the terms of the \CRF.
This process of verbal command updating cost, followed by automatic inference to get the results,
is repeated until satisfactory results are achieved.
Putting the human in the loop allows one to quickly obtain very good results.
This is because the user can intuitively leverage a high level understanding
of the current image and quickly find discriminative visual attributes to improve scene parsing.
For example, in \figref{fig:teaser}(c), if the verbal command contains the words `glass picture',
our algorithm can re-weight \CRF~to allow improved parsing of the `picture' and the `glass'.
Finally, we show that our joint
\CRF~formulation can be factorized. 
This permits the use of efficient filtering based techniques
\cite{krahenbuhl2011efficient}
to perform inference at interactive speed.

We evaluate our approach on the attribute-augmented \NYU~V2 RGB image
dataset~\cite{silberman2012indoor} that contains 1449 indoor images.
We compare our results with \sArt object-based image parsing
algorithms~\cite{AssociativeHCRFICCV2009,krahenbuhl2011efficient}.
We report a $6\%$ improvement in terms of average label accuracy ({\sc ala})
\footnote{Label accuracy is defined as the number of pixels with correct label
divided by the total number of pixels.}
using our automated object/attribute image parsing.
Beyond these numbers, our algorithm provides critical verbal handles for refinement and
subsequent edits leading to a significant improvement ($30\%$ {\sc ala})
when verbal interaction is allowed.
Empirically, we find that our interactive joint image parsing results
are better aligned with human perception than those of previous non-interactive approaches,
as validated by extensive evaluation results provided in the supplementary material.
Further, we find our method performs well on similar scene types
taken from outside of our training database.
For example, our indoor scene parsing system works on internet images
downloaded using `bedroom' as a search word in Google.

Whilst scene parsing is important in its own right,
we believe that our system enables novel human-computer  interactions.
Specifically, by providing a hands-free selection mechanism
to indicate objects of interest to the computer,
we can largely replace the role traditionally filled by the mouse.
This enables interesting image editing modalities such
as verbal guided image manipulation which can be integrated in smart phones and
Google Glass, by making commands such as  `zoom in on the cupboard in the far right'
meaningful to the computer.

In summary, our main contributions are:
\begin{itemize}
  \item a new interaction modality that enables verbal
    commands to guide image parsing;
  \item the development of a novel multi-label factorial \CRF~that can
    integrate cues from multiple sources at interactive rates; and
  \item a demonstration of the potential of this approach to make
    conventional mouse-based tasks hands-free.
\end{itemize}

\section{Related works}

\myPara{Object class image segmentation and visual attributes}
Assigning an object label to each image pixel,
known as object class image segmentation or scene parsing,
is one of computer vision's core problems.
TextonBoost \cite{shotton2009textonboost},
is a ground breaking work for addressing this problem.
It simultaneously achieves pixel-level object class recognition and segmentation
by jointly modeling patterns of texture and their spatial layout.
Several refinements of this method have been proposed,
including context information modeling~\cite{rabinovich2007objects},
joint optimization of stereo and object label \cite{LuborBMVC2010},
dealing with partial labeling \cite{verbeek2007scene},
and efficient inference~\cite{krahenbuhl2011efficient}.
These methods  deal only with object labels (noun) and not attributes (adjectives).
Visual attributes~\cite{VisualAttributesNIPS2007} and data association \cite{malisiewiczCvpr08},
which describe important semantic properties of objects,
have been shown to be an important factor for improving object recognition
\cite{farhadi2009describing,wang2010discriminative},
scene attributes classification \cite{SunAttributesDataCVPR2012},
and even modeling of unseen objects \cite{LearningtoDetectUnseenCVPR2009}.
These works have been limited to determining the
attributes of an image region contained in a rectangular bounding box.
Recently, Tighe and Lazebnik~\shortcite{understandingscenesICCV2011} have
addressed the problem of parsing image regions with multiple label sets.
However, their inference formulation remains unaware of object boundaries
and the obtained object labeling usually spreads over the entire image.
We would like to tackle the problem of image parsing with both objects and attributes.
This is a very difficult problem as, in contrast to traditional
image parsing in which only one label is predicted per pixel,
there now might be zero, one, or a set of labels predicted for each pixel,
e.g. a pixel might belong to wood, brown, cabinet, and shiny.
Our model is defined on pixels with fully connected graph topology,
which has been shown~\cite{krahenbuhl2011efficient} to be able to produce fine detailed boundaries.

\myPara{Interactive image labeling}
Interactive image labeling is an active research field. This field has two distinct trends.
The first involves having some user defined scribbles or bounding boxes,
which are used to assist the computer in cutting out the desired object from image
\cite{PaintSelection2009,LazySnapping2004,rother2004grabcut,imgsegwithbdx2009}.
Gaussian mixture models (\GMM) are often employed to model the color distribution of foreground
and background.
Final results are achieved via  Graph Cut~\cite{graphcut2001}.
While widely used, these works do not extend naturally to verbal parsing as
the more direct scribbles cannot be replaced with vague verbal descriptions such as `glass'.
The second trend in interactive image labeling incorporates a human-in-the-loop
\cite{BransonhumansintheloopECCV2010,humanintheloopiccv2011},
which focuses on recognition of image objects rather than image parsing.
They resolve ambiguities by interactively asking users to click on the object parts
and answer yes/no questions. 
Our work can be considered a verbal guided human-in-the-loop image parsing.
However, our problem is more difficult than the usual human-in-the-loop problems
because of the ambiguity of words (as opposed to binary answers to questions)
and the requirement for fine pixel wise labeling (as opposed to categorization).
This precludes usage of a simple tree structure for querying and  motivates our
more sophisticated,
interactive joint \CRF~model  to  resolve the ambiguities.

\myPara{Semantic-based region selection} Manipulation in the semantic space \cite{berthouzoz2011framework}
is a powerful tool and there are a number of approaches. An example is
Photo Clip Art \cite{lalonde2007photo} which allows users to directly
insert new semantic objects into existing images,
by retrieving suitable objects from a database.
This work has been further extended to sketch based
image composition by automatically extracting and selecting suitable
salient object candidates \cite{11cvprChengSaliency}
from Internet images \cite{ChenCT09Sketch2Photo,ChengPAMI,goldberg2012data}.
Carroll \etal \shortcite{carroll2010image} enables perspective aware image warps by using
user annotated lines as projective constrains.
Cheng \etal~\shortcite{ChengZMHH10} analyze semantic object regions
as well as layer relations according to user input scribble marking,
enabling interesting interactions across repeating elements.
Zhou \etal \shortcite{zhou2010parametric} proposed to reshape human image regions
by fitting an appropriate $3${\sc d} human model.
Zheng \etal \shortcite{zcczhm_interactiveImages_sigg12} partially recover the $3${\sc d} of
man-made environments, enabling intuitive non-local editing.
However, none of these methods attempt interactive verbal guided image parsing
which has the added difficulty of enabling
the use of verbal commands to provide vague guidance cues.

\myPara{Speech interface} Speech interfaces are deployed when mouse based interactions
are infeasible or cumbersome.
Although research on integrating speech interfaces into software started in the 1980s \cite{bolt1980put},
it is only recently that such interfaces have been widely deployed,
(e.g. Apple's Siri, PixelTone~\shortcite{laput2013pixeltone}).
However, most speech interface research is focused on natural language processing and
to our knowledge there has been no prior work addressing image region selection through speech.
The speech interface that most resembles our work is PixelTone
\shortcite{laput2013pixeltone},  which  allows users to attach object labels to
scribble based segments. These labels allow subsequent voice reference.
Independently, we have developed a hands-free parsing of an image
into pixel-wise object/attribute labels that correspond to human semantics.
This provides a verbal option for selecting objects of interest and is potentially,
a powerful additional tool for speech interfaces.

\begin{algorithm}[b!]
    \begin{algorithmic}
        \STATE \textbf{Input:} an image and object/attributes potentials (see \figref{fig:teaser}).
        \STATE \textbf{Output:} an object and a set of attributes labels for each pixel.
        \STATE \textbf{Initialize:} object/attributes potentials for each pixel; find pairwise potentials by \equref{eq:Pairwise}.
        \FOR{Automatic inference iterations $i$ = 1 to $T_{a}$}
            \STATE Update potentials using \equref{eq:update1} and \equref{eq:update2} for all pixels simultaneously using efficient filtering technique;
        \ENDFOR
        \FOR{each verbal input}
            \STATE update potentials (\cf \secref{sec::Interaction}) according to user input;
            \FOR{Verbal interaction iterations $i$ = 1 to $T_{v}$}
                \STATE Update potentials using \equref{eq:update1} and \equref{eq:update2}  as before;
            \ENDFOR
        \ENDFOR
        \STATE \textbf{Extract results from potentials:} at any stage, labels for each pixel could be found by selecting the largest object potential, or comparing the positive and negative attributes potentials.
    \end{algorithmic}
    \caption{Verbal guided image parsing.
    }\label{alg:Model}
\end{algorithm}

\section{System Design}

\begin{figure}[t!]
  \centering
  \includegraphics[width=.95\linewidth]{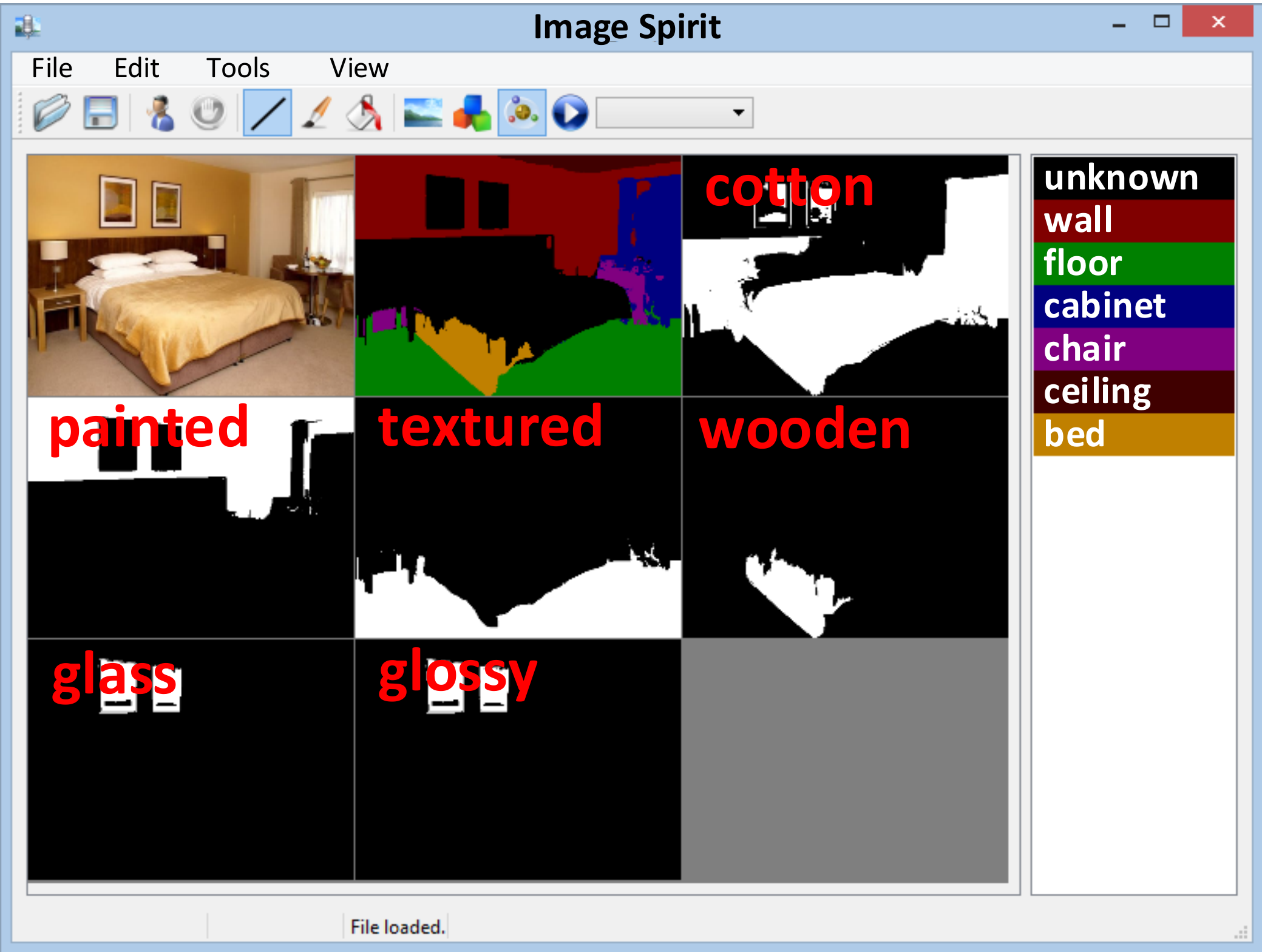}\\ \vspace{-0.08in}
  \caption{User interface of our system (labeling thumbnail view).
  }\label{fig:ui}
\end{figure}

Our goal is a verbal guided image parsing system that is simple, fast,
and most importantly, intuitive,
i.e. allowing an interaction mode  similar to our everyday language.
After the user loads an image, our system automatically assigns an object
class label (noun) and sets of attribute labels (adjectives) to each pixel.
Based on the initial automatical image parsing results, our system identifies a subset of objects and attributes
that are most related to the image.
In \figref{fig:ui}, 
to speed up the inference in the verbal refinement stage,
our system only consider the subset instead of the whole set of object classes and the attribute labels.
The initial automatic image parsing results also provide the bridge between image pixels and verbal commands.
Given the parse, the user can use his/her knowledge about the image
to strengthen or weaken various object and attribute classes.
For example, the initial results in \figref{fig:ui} might prompt the user to
realize that the bed is missing from the segmentation but the `cotton' attribute
covers a lot of the same area as is covered by the bed in the image.
Thus, the simple command
`Refine the cotton bed in center-middle' will strengthen the association between
cotton and bed, allowing a better segmentation of the bed.
Note that the final object boundary does not necessarily follow
the original boundary of the attribute because  verbal information is incorporated
only as soft cues, which are interpreted by a \CRF~within
the context of the other information.
\algref{alg:Model} presents a high level summary of our verbal guided
image parsing pipeline,
with details explained in the rest of this section.

Once objects have been semantically segmented,
it becomes straightforward to manipulate them using verb-based commands such as move,
change, etc.
As a demonstration of this concept, we encapsulate a series of rule-based image processing
commands needed to execute an action, allowing hands-free image manipulation
(see \secref{sec:applications}).

\subsection{Mathematical Formulation}\label{sec::fmCRF}

We formulate simultaneous semantic image parsing for object class and
attributes as a multi-label \CRF~that encodes both object and attribute classes,
and their mutual relations.
This is a combinatorially large  problem. If
each  pixel takes  one of the $16$ object labels and a subset of $8$ different attribute labels,
there are $(16 \times 2^{8})^{640\times480}$ possible solutions to consider for an image
of resolution $640 \times 480$.
Direct optimization over such a huge number of variables is computational
infeasible without some choice of simplification.
The problem becomes more complicated if correlation between attributes and
objects are taken into account.
In this paper, we propose using a factorial \CRF~framework \cite{sutton2004dynamic}
to model correlation between objects and attributes.

A multi-label \CRF~for dense image parsing of objects and attributes can be defined over
random variables $\mathcal{Z}$,
where each $Z_i = (X_i, Y_i)$ represents object and attributes variables
of the corresponding image pixel $i$ (see \tabref{tab:Anotations} for a list of notations).
$X_i$ will take a value from the set of object labels, $x_i \in \mathcal{O}$.
Rather than taking values directly in the set of attribute labels $\mathcal{A}$,
$Y_i$ takes values from the power-set of the attributes.
For example, $y_i = \{wood\}$, $y_i = \{wood, painted, textured\}$,
and $y_i=\emptyset$ are all valid assignments.
We denote by $\mathbf{z}$ a joint configuration of these random variables,
and $\mathbf{I}$ the observed image data.
Our \CRF~model is defined as the sum of per pixel and pair of pixel terms:
\begin{equation}\label{eq:joint}
E(\mathbf{z}) = \sum_{i} \psi_i (z_i) + \sum_{i < j} \psi_{ij} (z_i,z_j), \vspace{-.08in}
\end{equation}
where $i$ and $j$ are pixel indices that range from $1$ to $N$.
The per pixel term $\psi_i (z_i)$ measures the cost of assigning an object
label and a set of attributes label to pixel $i$,
considering learned pixel classifiers for both objects and attributes,
as well as learned object-attribute and attribute-attribute correlations.
The cost term $\psi_{ij} (z_i,z_j)$ encourages similar and nearby pixels
to take similar labels.

\begin{table}[!t]
    \centering
  \fontsize{8}{1.1em}\selectfont
    \begin{tabular}{c|l} \hline\hline
      Symbols &  Explanation (use RV to represent random variable)  \\ \hline
      $\mathcal{O}$ & Set of object labels: $\mathcal{O} = \{o_1,o_2,...,o_K\}$  \\ \hline
      $\mathcal{A}$ & Set of attribute labels: $\mathcal{A} = \{a_1,a_2,...,a_M\}$  \\ \hline
      $\mathcal{P(A)}$ & Power set of $\mathcal{A}$: $\mathcal{P(A)} = \{\{\}, \{a_1\}, ..., \{a_1, ..., a_M\}\}$\\ \hline
      $X_i$         & A RV for object label of pixel $i \in \{1, 2, ..., N\}$ \\ \hline
      $Y_{i,a}$     & A RV for attribute $a \in \mathcal{A}$ of pixel $i$  \\ \hline
      $Y_i$         & A RV for a set of attributes $\{a : Y_{i,a} = 1\}$ of pixel $i$ \\ \hline
      $Z_i$         & A RV $Z_i = (X_i, Y_i)$ of pixel $i$   \\ \hline
      $\mathcal{Z}$ & RVs of \CRF: $\mathcal{Z} = \{Z_1,Z_2,...,Z_N\}$  \\ \hline
      $y_{i,a}, y_i$& Assignment of RVs $Y_i, Y_{i,a}$: $y_{i,a} \in \{0, 1\}, y_i \in \mathcal{P(A)}$   \\ \hline
      $x_i, z_i$    & Assignment of RVs $X_i, Z_i$: $x_i \in \mathcal{O}$, $z_i = (x_i, y_i)$ \\ \hline
      $\psi_i$      & Unary cost of \CRF~ \\ \hline
      $\psi_{ij}$   & Pairwise cost of \CRF~ \\ \hline
      $\psi_i^{\mathcal{O}}(x_i)$           & Cost of $X_i$ taking value $x_i \in \mathcal{O}$  \\ \hline
      $\psi_{i,a}^{\mathcal{A}}(y_{i,a})$   & Cost of $Y_{i,a}$ taking value $y_{i,a} \in \{0, 1\}$  \\ \hline
      $\psi_{i,o,a}^{\mathcal{OA}}$         & Cost of conflicts between correlated attributes and objects  \\ \hline
      $\psi_{i,a,a'}^{\mathcal{A}}$         & Cost of correlated attributes taking distinct indicators  \\ \hline
      $\psi_{ij}^{\mathcal{O}}$             & Cost of similar pixels with distinct object labels  \\ \hline
      $\psi^{\mathcal{A}}_{i,j,a}$          & Cost of similar pixels with distinct attribute labels  \\ \hline
    \end{tabular}
    \tbl{List of notations}
    \label{tab:Anotations}
\end{table}

\begin{figure}[t]
  \centering
  \includegraphics[width=\linewidth]{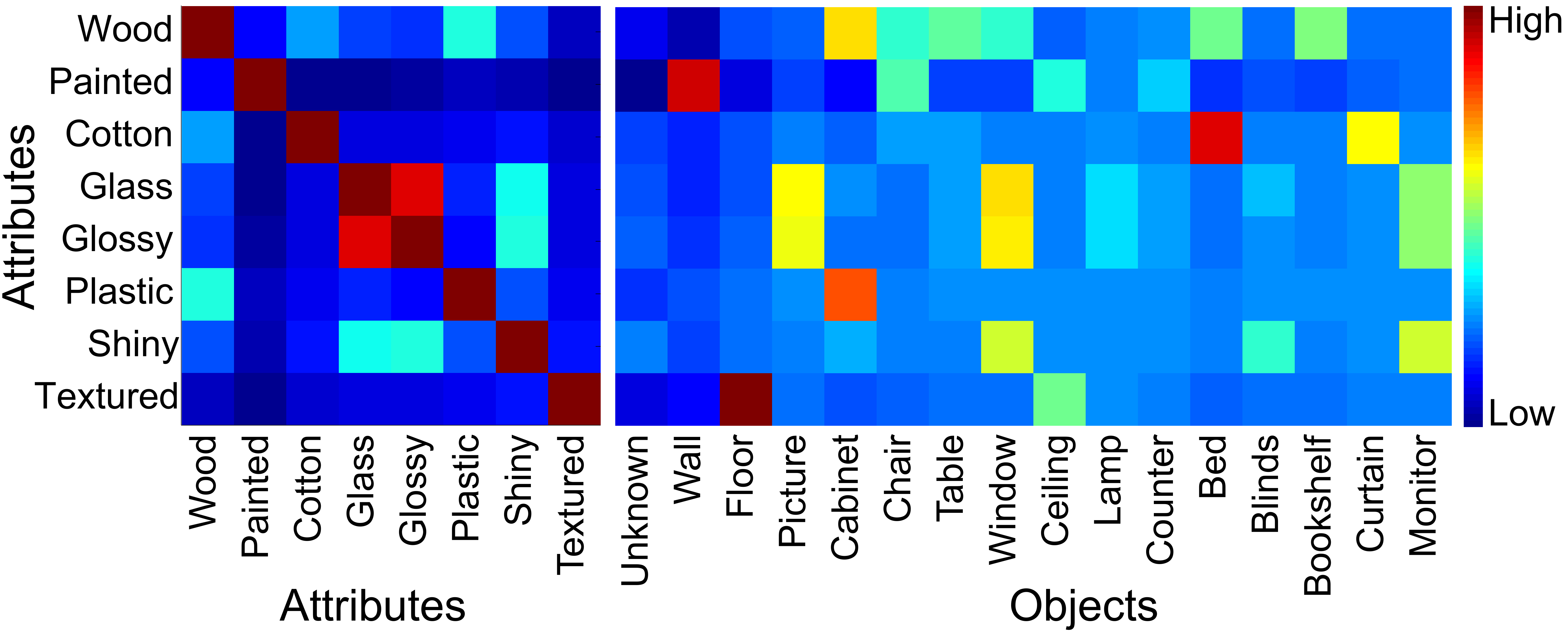}\\ \vspace{-.08in}
  \caption{Visualization of the $R^\mathcal{OA}, R^\mathcal{AA}$ terms used
    to encode object-attribute and attribute-attribute relationships.
  }\label{fig:corelations}
\end{figure}

To optimize \equref{eq:joint} we break it down into multi-class and binary
subproblems using a factorial \CRF~framework \cite{sutton2004dynamic},
whilst maintaining correlations between object and attributes.
The pixel term is decomposed into:
\begin{eqnarray}\label{eq:jointFull}
 \psi_i(z_i) & = & \psi_i^{\mathcal{O}}(x_i) + \sum_a \psi_{i,a}^{\mathcal{A}}(y_{i,a})
                 + \sum_{o,a} \psi_{i,o,a}^{\mathcal{OA}}(x_i,y_{i,a}) \nonumber \\
             &  & + \sum_{a \neq a'} \psi_{i,a,a'}^{\mathcal{A}}(y_{i,a},y_{i,a'})
\end{eqnarray}
where the cost of pixel $i$ taking object label $x_i$ is
$\psi_i^{\mathcal{O}}(x_i) = -log(\mbox{Pr}(x_i))$,
with probability derived from trained pixel classifier (TextonBoost~\cite{shotton2009textonboost}).
For each of the $M$ attributes, we train independent binary TextonBoost classifiers,
and set $\psi_{i,a}^{\mathcal{A}}(y_{i,a}) = -log(\mbox{Pr}(y_{i,a}))$
based on the output of this classifier.
Finally, the terms $\psi_{i,o,a}^{\mathcal{OA}}(x_i,y_{i,a})$
and $\psi_{i,a,a'}^{\mathcal{A}}(y_{i,a},y_{i,a'})$
are the costs of correlated objects and attributes with distinct indicators.
They are defined as:
\begin{eqnarray}\label{eq:corelated}
    \psi_{i,o,a}^{\mathcal{OA}}(x_i,y_{i,a}) =
    [[x_i=o] \neq y_{i,a}] \cdot \lambda_{\mathcal{OA}}R^\mathcal{OA}(o,a) \nonumber \\
    \psi_{i,a,a'}^{\mathcal{A}}(y_{i,a},y_{i,a'}) =
    [y_{i,a} \neq y_{i,a'}] \cdot \lambda_{\mathcal{A}}R^\mathcal{A}(a,a')
\end{eqnarray}
where Iverson bracket, $[.]$, is 1 for a true condition and 0 otherwise,
$R^\mathcal{OA}(o,a)$ and and $R^\mathcal{A}(a,a')$ are derived from learned
object-attribute and attribute-attribute correlations respectively.
 Here $\psi_{i,o,a}^{\mathcal{OA}}(x_i,y_{i,a})$ and
$\psi_{i,a,a'}^{\mathcal{A}}(y_{i,a},y_{i,a'})$ penalize inconsistent
object-attributes and attribute-attribute labels by the cost of
their correlation value.
These correlations are obtained from the $\phi$ coefficient,
which is learnt from the labeled dataset using \cite{Correlation2BRapproach2009}.
A visual representation of these correlations is given in \figref{fig:corelations}.

The cost term $\psi_{ij} (z_i,z_j)$ can be factorized as object label
consistency term and attributes label consistency terms:
\begin{equation}
\label{eq:Pairwise}
\psi_{ij}(z_i,z_j) = \psi_{ij}^{\mathcal{O}}(x_i,x_j) +
        \sum_a \psi^{\mathcal{A}}_{i,j,a}(y_{i,a},y_{j,a}),
\end{equation}
here we assume each has the form of Potts model \cite{potts1952some}:\\ \vspace{-.1in}
\begin{eqnarray}
    \psi^{\mathcal{O}}_{ij}(x_i,x_j) = [x_i \neq x_j] \cdot g(i,j)  \nonumber \\
    \psi^{\mathcal{A}}_{i,j,a}(y_{i,a},y_{j,a}) = [y_{i,a} \neq y_{j,a}] \cdot g(i,j). \nonumber
\end{eqnarray}
We define $g(i,j)$ in terms of similarity between color vectors
$I_i$, $I_j$ and position values $p_i$, $p_j$:
\begin{eqnarray}
\label{eq:PairSimilarity}
g(i,j) & = & w_1 \exp(-\frac{|p_i - p_j|^2}{2\theta_{\mu}^{2}} -
\frac{|I_i - I_j|^2}{2\theta_{\nu}^{2}}) \nonumber \\
& & + w_2 \exp(- \frac{|p_i-p_j|^2}{2\theta_{\gamma}^2}).
\end{eqnarray}
All the parameters $\lambda_{\mathcal{OA}}$, $\lambda_{\mathcal{A}}$, $w_1$, $w_2$,
$\theta_{\mu}$, $\theta_{\nu}$, and $\theta_{\gamma}$ are learnt via cross validation.
%

\subsection{Efficient Joint Inference with Factorized Potentials}


To enable continuous user interaction,
our system must have a response rate which is close to real time.
Recently there has been a breakthrough in the mean-field solution of random fields,
based on advances in filtering based methods in computer graphics~\cite{Adams_fasthigh-dimensional,krahenbuhl2011efficient}.
Here we briefly sketch how this inference can be extended to multi label \CRF s.

This involves finding a mean-field approximation $Q(\mathbf{z})$
 of the true distribution $P \propto \exp(-E(z))$,
by minimizing the KL-divergence $D(Q||P)$ among all distributions $Q$
that can be expressed as a product of independent marginals,
$Q(\mathbf{z}) = \prod_{i} Q_{i} (z_i)$.
Given the form of our factorial model,
we can factorize $Q$ further into a product of marginals
over multi-class object and binary attribute variables.
Hence we take $Q_i(z_i) = Q^{\mathcal{O}}_{i} (x_i)\prod_{a}Q^{\mathcal{A}}_{i,a} (y_{i,a})$,
where $Q^{\mathcal{O}}_{i}$ is a multi-class distribution over the object labels,
and $Q^{\mathcal{A}}_{i,a}$ is a binary distribution over $\{0,1\}$.

Given this factorization,
we can express the required mean-field updates
(\cf \cite{koller2009probabilistic}) as:
\begin{align}\label{eq:update1}
Q_i^{\mathcal{O}}(x_i = o) = & \frac{1}{Z^{\mathcal{O}}_i}\exp\{-\psi^{\mathcal{O}}_i(x_i) \nonumber \\
  &  - \sum_{i \neq j}Q_j^{\mathcal{O}}(x_j=o)(-g(i,j)) \nonumber \\
  & - \sum_{a \in \mathcal{A}, b\in\{0,1\}}Q_{i,a}^{\mathcal{A}}(y_{i,a}=b)\psi^{\mathcal{OA}}_{i,o,a}(o,b)\}
\end{align}
\begin{align}\label{eq:update2}
Q_{i,a}^{\mathcal{A}}(y_{i,a}=b) =& \frac{1}{Z^{\mathcal{A}}_{i,a}}\exp\{-\psi^{\mathcal{A}}_{i,a}(y_{i,a})  \nonumber \\
&  -\sum_{i \neq j}Q_{j,a}^{\mathcal{A}}(y_{j,a}=b)(-g(i,j))  \nonumber \\
& -\sum_{a^{\prime}\neq a \in \mathcal{A}, b^{\prime}\in\{0,1\}}Q_{i,a'}^{\mathcal{A}}(y_{i,a'}=b^{\prime})
    \psi^{\mathcal{A}}_{i,a,a^{\prime}}(b, b^{\prime}) \nonumber \\
& - \sum_{o} Q_{i}^{\mathcal{O}}(x_{i}=o)\psi^{OA}_{i,o,a}(o,b)\}
\end{align}
where $Z^{\mathcal{O}}_i$ and $Z^{\mathcal{A}}_{ia}$ are per-pixel object and attributes normalization factors.
As shown in \equref{eq:update1} and \equref{eq:update2},
directly applying these updates for all pixels requires expensive sum operations,
whose computational complexity is quadratic in the number of pixels.
Given that our pair of pixel terms are of Potts form modulated by a linear combination of Gaussian kernels
as described in \equref{eq:PairSimilarity}, simultaneously finding these sums for all pixels can be achieved at a complexity
linear in the number of pixels using efficient filtering techniques \cite{Adams_fasthigh-dimensional,krahenbuhl2011efficient}.

\begin{figure}
    \fontsize{8}{10}
    \fbox{\parbox{.97\linewidth}{%
    \begin{algorithmic}
        \STATE \hspace{-.15in} Basic definitions:
        \STATE {\bf MA}, {\bf SA}, {\bf CA}, {\bf PA}, are attributes keywords in \secref{sec::Interaction}.
        \STATE {\bf Obj} is an object class name keyword in \secref{sec::Interaction}.
        \STATE {\bf ObjDes} := [{\bf CA}] [{\bf SA}] [{\bf MA}] {\bf Obj} [in {\bf PA}]
        \STATE {\bf DeformType} $\in$ \{`lower', `taller', `smaller', `larger'\}
        \STATE {\bf MoveType} $\in$ \{`down', `up', `left', `right'\}
        \STATE
        \STATE \hspace{-.15in} Verbal commands for image parsing:
        \STATE Refine the {\bf ObjDes}.
        \STATE Correct the {\bf ObjDes} as {\bf Obj}.
        \STATE \hspace{-.15in} Verbal commands for manipulation:
        \STATE Activate the {\bf ObjDes}.
        \STATE Make the {\bf ObjDes} {\bf DeformType}.
        \STATE Move the {\bf ObjDes} {\bf MoveType}.
        \STATE Repeat the {\bf ObjDes} and move {\bf MoveType}.
        \STATE Change the {\bf ObjDes} [from {\bf Material/Color}] to {\bf Material/Color}.
    \end{algorithmic}
    }}
    \caption{Illustration of supported verbal commands for image
        parsing and manipulation.
        The brackets `[]' represent optional words.}
    \label{fig:commands} \vspace{-.05in}
\end{figure}

\subsection{Refine Image Parsing with Verbal Interaction}
\label{sec::Interaction}

Since the image parsing results of the automatic approach described
in~\secref{sec::fmCRF} are still far away from what is our human being perceive
from the image and what is required by most image parsing applications such as
photo editing, we introduce a verbal interaction modality so that the user can
refine the automatic image parsing results by providing a few verbal commands.
Each command will alter one of the potentials given in~\secref{sec::fmCRF}.

Supported object classes  (\textbf{Obj})  include the
16 keywords in our training object class list (bed, blinds, bookshelf,
cabinet, ceiling, chair, counter, curtain, floor, lamp, monitor, picture,
table, wall, window and unknown).
We also support 4 material attributes (\textbf{MA}) keywords
(wooden, cotton, glass, plastic)
and 4 surface attributes  (\textbf{SA})  keywords (painted, textured, glossy, shiny).
For color attributes  (\textbf{CA}), we support the 11 basic color names,
suggested by Linguistic study \cite{berlin1991basic}.
These colors names/attributes are:
black, blue, brown, grey, green, orange, pink, purple, red, white and yellow.
Also as observed by \cite{laput2013pixeltone}, humans are not good at describing
precise locations but can easily refer to some rough positions in the image.
We currently support $9$ rough positional attributes (\textbf{PA}),
by combining $3$ vertical positions (top, center, and bottom)
and $3$ horizontal positions (left, middle, and right).

\figref{fig:commands} illustrates the $7$ commands that are currently supported.
These command can alter the per pixel terms in \equref{eq:jointFull}.
Notice that both the image parsing commands (e.g. \tabref{tab:VerbalCommands})
and the manipulation commands (e.g. \figref{fig:Applications})
contain object descriptions (\textbf{ObjDes}) for verbal refinement.
If needed\footnote{When we have perfect image parsing results for the image to be manipulated,
we might verbally switch off the function that conducts this combination operation
of image parsing and manipulation.}, this enables the image parsing to be
updated during a manipulation operation.
In \figref{fig:commands} the distinction between commands `refine'
and `correct' is as follows: the former should be given when the label assignment is good
but the segment could be better; while,
the later is to be given when the label is incorrect.

Consider that user give verbal command `Refine the \textbf{ObjDes}', where
\textbf{ObjDes}=[\textbf{CA}][\textbf{SA}][\textbf{MA}]\textbf{Obj}[in \textbf{PA}]'.
The system understands there should be a object named \textbf{Obj} in the position \textbf{PA}, and the correlation cues such as \textbf{MA}-\textbf{SA}, \textbf{MA}-\textbf{Obj} and \textbf{SA}-\textbf{Obj} should be encouraged.
We achieve this by updating the correlation matrices  given in \equref{eq:corelated}.
Thus, the altered object-attribute correlations are changed as ${R'}^{\mathcal{OA}} = \lambda_1 +\lambda_2  R^{\mathcal{OA}}$
and the modified attribute-attribute correlations are updated as ${R'}^{\mathcal{A}}=\lambda_3 +\lambda_4  R^{\mathcal{A}}$
where $\lambda_i$ are tuning parameters.

\myPara{Speech parsing}
We use the freely available Microsoft speech {\sc sdk} \shortcite{speechSDK}
to convert a spoken command into text.
We use a simple speech grammar, with a small number of fixed commands.
Since the structure of our verbal commands and the candidate keywords list are fixed,
the grammar definition API of Microsoft speech SDK allows us to robustly capture
user speech commands.
For  more sophisticated speech recognition and parsing,
see \cite{laput2013pixeltone}.

\begin{figure}
    \centering
    \subfigure[source image]{\includegraphics[width=.325\linewidth]{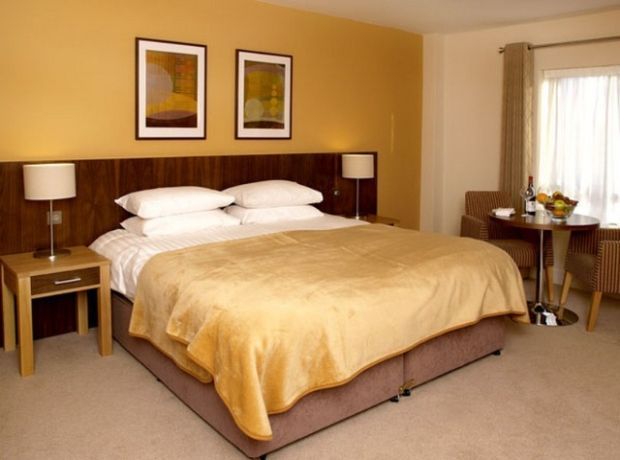}}
    \subfigure[white $\mathbf{R}_c$]{\includegraphics[width=.325\linewidth]{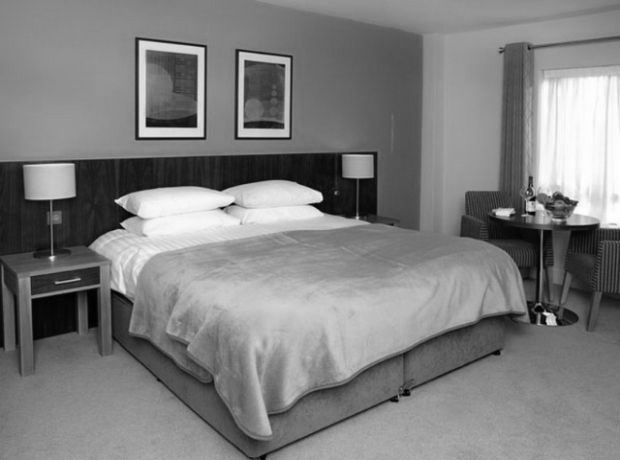}}
    \subfigure[center-middle $\mathbf{R}_s$]{\includegraphics[width=.325\linewidth]{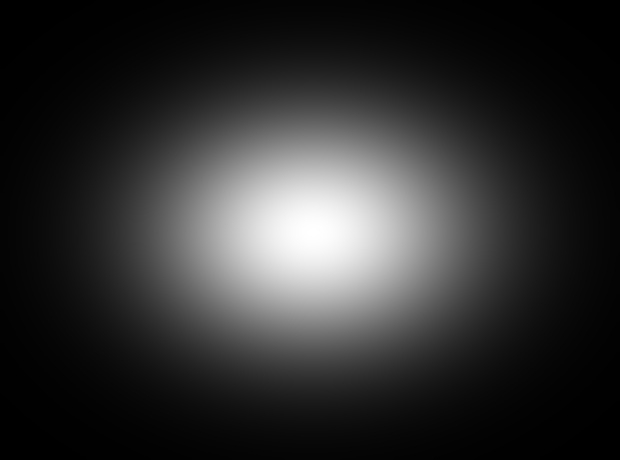}}\\
    \vspace{-.15in}
    \caption{Response maps of $\mathbf{R}_c$ and $\mathbf{R}_s$ for attributes `white' and `center-middle' respectively.
    }\label{fig:ColrSpatiAtri}
\end{figure}

\myPara{Color $\mathbf{R}_c$ and spatial $\mathbf{R}_s$ attributes response map}
Colors are  powerful attributes
that can significantly improve performance of object classification
\cite{van2010evaluating} and detection \cite{shahbaz2012color}.
To incorporate color into our system, we create a color response map,
with the value at the $i$th pixel defined according to the distance
between the color of this pixel $I_i$ and a user specified color $\mathbb{I}$.
We use $R_c(i) = 1 - \|I_i - \mathbb{I}\|$,  where each of the {\sc rgb}
color channels are in the range [0,1].
We also utilize the location information present in the command to localize objects.
Similar to color, the spatial response map value at the $i$th pixel
is defined as $R_s(i) = \exp(-\frac{d^2}{2\delta^2})$,
where $d$ is the distance between the pixel location and the user indicated position.
In the implementation, we use $\delta^2 = 0.04$ with pixel coordinates in both directions
normalized to [0,1].
\figref{fig:ColrSpatiAtri} illustrates an example of color
and position attributes generated according to a given verbal command.
The spatial and color response maps are combined into a final overall map
$R(i)=R_s(i)R_c(i)$ that is used to update per pixel terms in \equref{update}.
Since rough color and position names are typically quite inaccurate,
we average the initial response values  within each region generated
by the unsupervised segmentation method \cite{felzenszwalb2004efficient}
for better robustness.
These response maps are normalized to the same range as other object classes'
per pixel terms for comparable influence to the learned object per pixel terms.

We use these response maps to update the corresponding
object and attribute per pixel terms, $\psi_i^{\mathcal{O}}(x_i),  \psi_{i,a}^{\mathcal{A}}(y_{i,a})$ in \equref{eq:jointFull}.
Specifically, we set
\begin{equation}
\label{update}
{\psi'}^{\mathcal{O}}_i(x_i) =
\psi_i^{\mathcal{O}}(.)- \lambda_5 R(i),\mbox{ if } x_i=\mathbb{O}
\end{equation}
where $\psi^{\mathcal{O}}_i(x_i)$ is the per pixel term for objects
and $\mathbb{O}$ is the user specified object.
Attribute terms are updated in a similar manner and share the same $\lambda_5$ parameter.
The $\lambda_{1, ..,5}$ parameters are set via cross validation.
After these per pixel terms are reset, the inference is re-computed to obtain the updated image parsing result.

\myPara{Working set selection for efficient interaction}
Our \CRF~is factorized for efficient inference over the full
set of object and attribute labels.
However, since the time it takes to perform inference is dependent on the number
of labels that are considered,
the interaction may 
take much longer if there are many labels.
To overcome this problem, a smaller working set of labels
can be employed during interaction, guaranteeing a smooth user experience.
Moreover, as observed in \cite{SturgessLCT12},
the actual number of object classes present in an image,
is usually much smaller than the total number of object-classes considered
(around a maximum of 8 out of 397  in the {\sc sun} database \cite{xiao2010sun}).
We exploit this observation by deriving the working set
as the set of labels in the result of our automatic parsing parse
and then updating it as required during interaction,
for instance if the user mentions a label currently not in the subset.
In our implementation this strategy gives an average timing of around 0.2-0.3
seconds per interaction, independent of the total number of labels considered.


\section{Evaluation}

\begin{figure}[h]
  \centering
  \includegraphics[width=\linewidth]{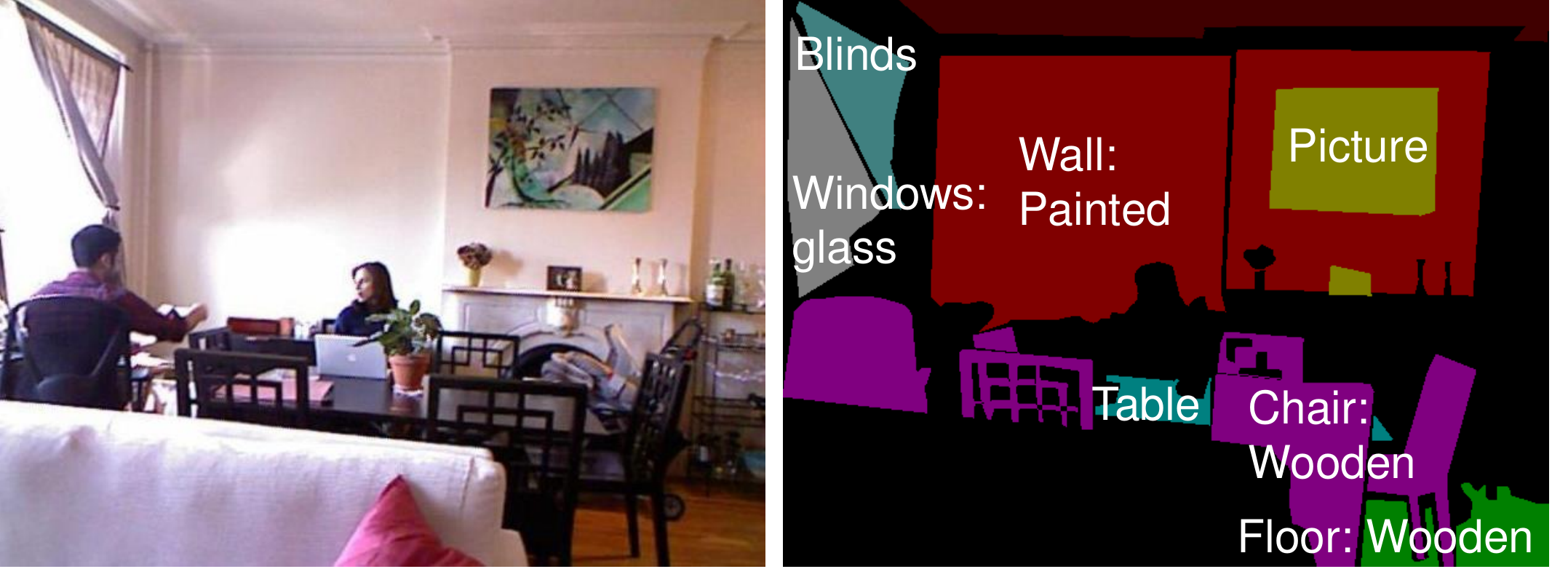}\\ \vspace{-.1in}
  \caption{Example of ground truth labeling in a\NYU~dataset: original image (left)
    and object class and attributes labeling (right).
  }\label{fig:aNYU}
\end{figure}

\myPara{a\NYU~Dataset (attributes augmented \NYU)}
We created a dataset for our evaluation since
per-pixel joint object and attributes segmentation
is an emerging problem thus there are only a few existing benchmarks\footnote{
As also noted by \cite{understandingscenesICCV2011},
although the \CORE~dataset \cite{AttributeCrossCVPR2010} contains object and attributes labels,
each \CORE~image only contains a single foreground object,
without background annotations.}.
In order to train our model and perform quantitative evaluation,
we augment the widely used {\sc nyu} indoor V2 dataset ~\cite{silberman2012indoor},
through additional manual labeling of  semantic attributes.
\figref{fig:aNYU} illustrates an example of  ground truth labeling of this dataset.
We use the \NYU~images with ground truth object class labeling,
and split the dataset into $724$ training images and $725$ testing images.
The list of object classes and attributes we use can be found in \secref{sec::Interaction}.
We only use the {\sc rgb} images from the \NYU~dataset although it provides
depth images.
Notice that each pixels in the ground truth images are marked with
an object class label and a set of attributes labels (on average, 64.7\% of them
are non empty sets).

\begin{figure*}
    \hfill
	\begin{overpic}[width=.985\textwidth,height=.92\textheight]{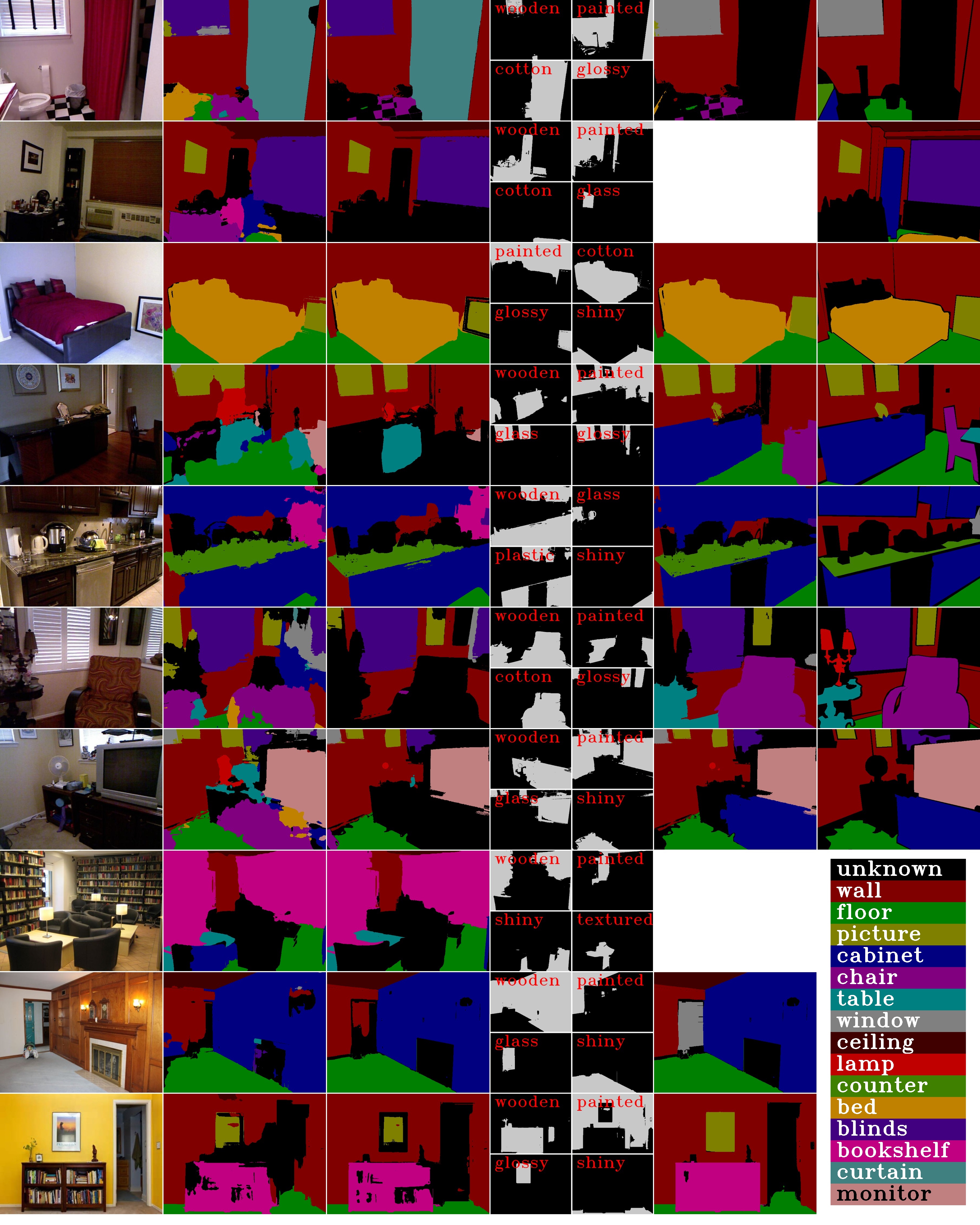}
        \put(1,-0.5){(a) source image}
        \put(17,-0.5){(b) DenseCRF}
        \put(30,-0.5){(c) our object class}
        \put(45,-0.5){(d) our attributes}
        \put(60,-0.5){(e) verbal refined}
        \put(74,-0.5){(f) ground-truth}
        \put(-2, 94){\rotatebox{90}{(1)}}
        \put(-2, 84){\rotatebox{90}{(2)}}
        \put(-2, 74){\rotatebox{90}{(3)}}
        \put(-2, 64){\rotatebox{90}{(4)}}
        \put(-2, 54){\rotatebox{90}{(5)}}
        \put(-2, 44){\rotatebox{90}{(6)}}
        \put(-2, 35){\rotatebox{90}{(7)}}
        \put(-2, 25){\rotatebox{90}{(8)}}
        \put(-2, 15){\rotatebox{90}{(9)}}
        \put(-2, 5){\rotatebox{90}{(10)}}
	\end{overpic}
	\caption{Qualitative comparisons. Note that after verbal refinement,
    our algorithm provides results that correspond closely to human scene understanding.
    This is also reflected in the numerical results tabulated in  \tabref{tab:AvgLabelAccuracy2}.
    The last three images are from the Internet and lack ground truth.
    For the second and eight image,
    there are no attribute combinations which would improve the result,
    hence there is no verbal refinement. (See \tabref{tab:VerbalCommands} for the used verbal commands.)
    \label{parsed}}
\end{figure*}

\myPara{Quantitative evaluation for automatic image parsing}
We conduct quantitative evaluation on a\NYU~ dataset.
Our approach consists of automatic joint objects-attributes image parsing and verbal guided image parsing.
We compared our approach against two state-of-the-art \CRF-based approaches including
Associative Hierarchical \CRF~approach~\cite{AssociativeHCRFICCV2009}
and Dense \CRF~\cite{krahenbuhl2011efficient}. For fair comparison, we train the same TextonBoost classifiers for all the methods
(a multi-class TextonBoost classifier for object class prediction and $M$
independent binary TextonBoost classifiers, one for each attributes).
Following \cite{krahenbuhl2011efficient},
we adopt the average label accuracy ({\sc ala}) measure for algorithm performance which is
the ratio between number of correctly labeled pixels and total number of pixels.
As shown in  \tabref{tab:AvgLabelAccuracy},
we have {\sc ala}~score of $56.6\%$
compared to $50.7\%$ for the previous \sArt results. During the
experiments, we achieve best results when we set $T_a = 5$, as described in \algref{alg:Model}.

\begin{table}
  \begin{tabular}{l|c|c|c|c} \hline\hline
      Methods        &~ H-CRF  &~ DenseCRF &~ Our-auto &~ Our-inter \\ \hline
      Label accuracy &  51.0\% & 50.7\%    & 56.9\%    & - -       \\ \hline
      Inference time &  13.2s  & 0.13s     & 0.54s     & 0.21s     \\ \hline
      Has attributes &    NO   & NO        & YES       & YES       \\ \hline \hline
  \end{tabular}
  \tbl{Quantitative results on  a\NYU~dataset.
    Qualitative results for all $725$ testing images can be found in the supplementary material.
    The H-CRF (Hierarchical conditional random field model) approach is implemented
    in a public available library: ALE(\protect\url{http://cms.brookes.ac.uk/staff/PhilipTorr/ale.htm}),
    Dense-CRF~\protect\cite{krahenbuhl2011efficient} represents the state-of-the-art \CRF~approach.
    Our-auto stands for our pixel-wise joint objects attributes image parsing approach.
    Our-inter means our verbal guided image parsing approach.
    All the experiments are carried out on a computer with Intel Xeon(E) $3.10$GHz CPU
    and $12$ GB RAM.
    Note that all methods in this table use the same features.
    Without the attributes terms,
    our \CRF~formulation will be reduced to exactly the same model as DenseCRF,
    showing that our JointCRF formulation benefits from the attributes components.
    Our-inter only considers the time used for updating the previous results given hints from user commands.
  }
  \label{tab:AvgLabelAccuracy}
\end{table}

\begin{table}
  \begin{tabular}{l|c|c|c} \hline\hline
      Methods     \hfill   & ~~~ DenseCRF ~~~ & ~~~ Our-auto ~~~ & ~~~ Our-inter ~~~ \\ \hline
      Label accuracy & 52.1\% & 56.2\%   & 80.6\%      \\ \hline \hline
  \end{tabular}
  \tbl{Evaluation for verbal guided image parsing.
    Here we show average statistics for interacting with a $50$ images subset.
  }\label{tab:AvgLabelAccuracy2}
\end{table}

\begin{table}[!b]
  \fontsize{8}{1.1em}\selectfont
  \begin{tabular}{c|l} \hline\hline
      Image & ~~~~~~~~~~~~~~~~~~~~~~~~~~~~~~~~~~~~~~~ Verbal commands   \\ \hline
      (1)  & Correct the blinds to window. Correct the curtain to unknown.  \\ \hline
      (3)  & Refine the glossy picture.   \\ \hline
      \Rows{(4)} & Refine the wooden cabinet in bottom-left. Refine the chair in \\
           & bottom-right. Refine the floor in bottom-middle. \\ \hline
      \Rows{(5)} & Refine the black plastic cabinet. Refine the white unknown in\\
           & bottom-middle. Refine the cabinet in bottom-left. \\ \hline
      \Rows{(6)} & Refine the cotton chair. Refine the glass unknown. Refine the  \\
           & black wooden table in bottom-left.   \\\hline
      (7)  & Refine the wooden cabinet in bottom-right.   \\ \hline
      (9)  & Refine the glass window.   \\ \hline
           & Refine the glossy picture. Refine the wooden bookshelf in \\
      (10) & bottom-middle. Refine the yellow painted wall in bottom middle. \\
           & Refine the textured floor. \\ \hline
      \hline
  \end{tabular}
  \tbl{Verbal commands used for parsing images in \figref{parsed}.}
  \label{tab:VerbalCommands}
\end{table}

\newcommand{\AddImg}[1]{\includegraphics[height=.1455\linewidth]{#1}}
\begin{figure*}[t]
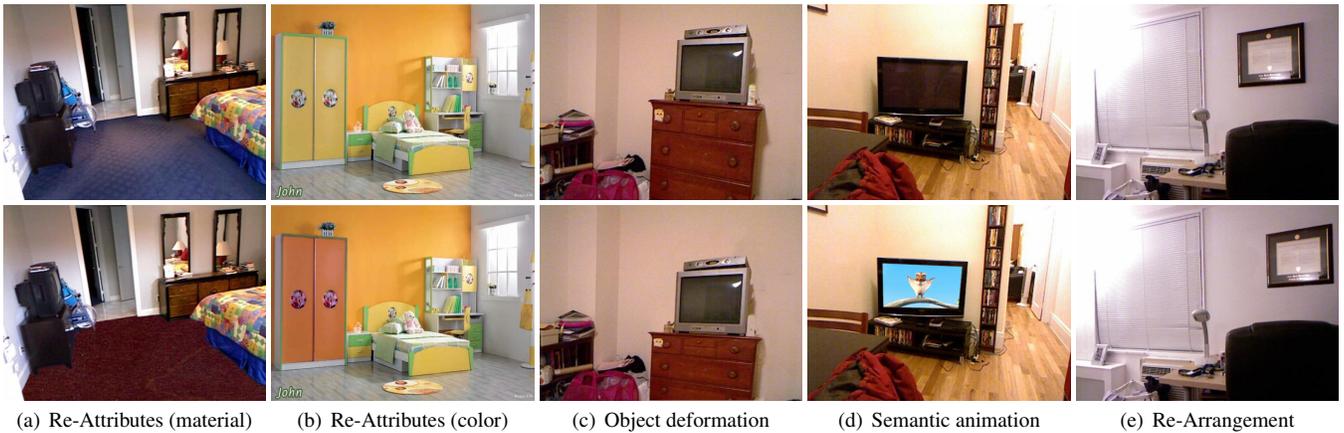

  \centering
  \AddImg{ReAttri/985.jpg}
  \AddImg{ReAttri/4.jpg}
  \AddImg{Deform/68.jpg}
  \AddImg{Animat/201.jpg}
  \AddImg{ReArrange/189.jpg}
  \\ \vspace{-.05in}
  \subfigure[Re-Attributes (material)]{\AddImg{ReAttri/985_Res.png}}
  \subfigure[Re-Attributes (color)]{\AddImg{ReAttri/4_Res.png}}
  \subfigure[Object deformation]{\AddImg{Deform/68_Res.png}}
  \subfigure[Semantic animation]{\AddImg{Animat/201_Res.png}}
  \subfigure[Re-Arrangement]{\AddImg{ReArrange/189_Res.png}}
  \\ \vspace{-.15in}
  \caption{Verbal guided image manipulation applications.
    The commands used are: (a) `Refine the white wall in bottom-left'
    and `Change the floor to wooden',
    (b) `Change the yellow wooden cabinet in center-left to brown',
    (c) `Refine the glossy monitor' and `Make the wooden cabinet lower',
    (d) `Activate the black shiny  monitor in center-middle',
    (e) `Move the picture right',
    See also the supplemental video.
  }\label{fig:Applications}
\end{figure*}

\myPara{Quantitative evaluation for verbal guided image parsing}
We numerically evaluate our verbal guided interaction.
We choose a subset of $50$ images whose collective accuracy
scores are reflective of the overall data set.
After verbal refinement, our accuracy rises to $80.6\%$ as compared to
the $50-56\%$ of automated methods.
From the results displayed in \figref{parsed},
one can see that these interactive improvements are not just numerical
but also produce object segmentations that accord more to human intuition.
In fact, many of the results appear similar to ground truth segments.
All evaluated images are shown in the supplementary material which bears out this trend.
In experiments, we achieve best speed-accuracy-tradeoff results when we set $T_a = 5$,
and $T_v = 3$, as described in \algref{alg:Model}.

Note that the final $3$ images of \figref{parsed} (more in supplementary)
are not part of the a\NYU~dataset but
are Internet images without any ground truth annotations.
These images demonstrate our algorithm's ability to generalize
training data for application to images from a similar class
(a system trained on indoor images will not work on outdoor scenes)
taken under uncontrolled circumstances.

\begin{table}
  \begin{tabular}{l|c|c|c} \hline\hline
      Interaction modality          & ~ verbal ~ & ~ touch ~ & verbal + touch \\ \hline
      Average interaction time (s)  &  6.6       &  32.3     & 11.7           \\ \hline
      Average accuracy (\%)         &  80.3      &  95.2     & 97.8           \\ \hline
      Average user preference (\%)  &  15.8      &  10.5     & 73.7           \\ \hline \hline
  \end{tabular}
  \tbl{Interactive time and accuracy comparison between different interaction
    modality: verbal, finger touch and both}
  \label{tab:UserStudy}
\end{table}

\myPara{User study}
Beyond large scale quantitative evaluation,
we also test the plausibility of our new interaction modality by a user study.
Our user study comprises of 38 participants, mostly computer science graduates.
We investigate both the time efficiency and the user preference of the verbal
interaction.
Each user was given a one page instruction script and 1 minute demo video to show
how to use verbal commands and mouse tools (line, brush,
and fill tool as shown in \figref{fig:ui})
to interact with the system.
The users were given 5 images and asked to improve the parsing results
using different interaction modality: i) only verbal, ii) only finger touch,
iii) both verbal and touch (in random order to reduce learning bias).
Statistics about average interaction time, label accuracy, and user preference
is shown in \tabref{tab:UserStudy}.
In our experiments, participants use a small number of
(mean and standard deviation: $1.6 \pm 0.95$) verbal commands
to roughly improve the automatic parsing results and then touch interaction
for further refinements.
In the `verbal+touch' modality, 74.4\% users preferred verbal command
before touch refinement.
In desktop setting, although average preference of verbal interaction is
not as good as touch interaction,
it provides a viable alternative to touch interaction
while the combination was generally preferred by most users.
We believe that for new generation devices such as Google Glass and other wearable devices,
our verbal interaction will be even more useful as it is not easy to perform traditional
interactions on them.

\section{Manipulation Applications}\label{sec:applications}

To demonstrate our verbal guided system's applicability as a selection mechanism,
we implement a hands-free image manipulation system.
After scene parsing has properly segmented the desired object,
we translate the verbs into pre-packaged sets of image manipulation commands.
These commands include in-painting \cite{sun2005image,09tog/BarnesSFG}
and alpha matting \cite{levin2008closed} needed  for a seamless editing effect,
as well as semantic rule-based considerations.
The list of commands supported by our system is given in \figref{fig:commands}
and some sample results in \figref{fig:Applications}.
The detailed effects are given below.
Although the hands-free image manipulation results are not entirely satisfactory,
we believe that the initial results demonstrate the possibility offered by verbal
scene parsing (see also video).

\myPara{Re-Attributes}
Attributes, such as color and  surface properties have a large impact on object appearance.
Changing these attributes is a common task and naturally lends itself to
verbal control. Once the scene has been parsed, one can verbally specify
the object to re-attribute.
As the computer has pixel-wise knowledge of the region the user is referring too,
it can apply the appropriate image processing operators to alter it.
Among all the pixels with user specified object class label,
we choose the 4-connected region with the biggest weight as the extent of the target object,
with weights defined by the response map as shown in \figref{fig:ColrSpatiAtri}.
Some examples are shown in  \figref{fig:Applications}.
To change object  color, we  add the difference
between average color of this object and the user specified target color.
For material changing, we simply tile  the target texture
(e.g. wood texture) within the object mask.
Alternately, texture transfer methods \cite{efros2001image} can be used.
Note that in the current implementation,
we ignore effects due to varying surface orientation. 

\myPara{Object Deformation and Re-Arrangement}
Once an object has been accurately identified, our system supports
move, size change and repeat commands that duplicate the object in a new region
or changes its shape. Inpainting is automatically carried out to refill exposed regions.
For robustness, we also define a simple, `gravity' rule for the `cabinet' and `table' classes.
This requires  small objects above these object segments
(except stuff such as wall and floor)
to  follow their motion.
Note that without whole image scene parsing,
this `gravity' rule is difficult to implement as there is a concern
that a background wall is defined as a small object.
Examples of these move commands can be seen in \figref{fig:Applications}c.

\myPara{Semantic Animation}
Real word objects often have their semantic functions.
For example, a monitor could be used to display videos.
Since we can estimate the object region and its semantic label,
a natural manipulation would be animating these objects by a set of user or predefined animations.
Our system supports an `activate' command.
By way of example consider \figref{fig:Applications},
when the user says `Activate the black shiny monitor in center-middle',
our system automatically fits the monitor region with a rectangle shape,
and shows a video in an detected inner rectangle  of the full monitor boundary
(typically related to screen area).
This allows the mimicking real world function of the monitor class.

\section{Discussion}

This paper presents a novel multi-label \CRF~formulation for efficient,
image parsing into per-pixel object and attribute labels.
The attribute labels act as verbal handles through
which users can control the \CRF,
allowing verbal refinement of the image parsing.
Despite the ambiguity of verbal descriptors,
our system can deliver fairly good image parsing results that correspond to human intuition.
Such hands-free parsing of an image provides verbal methods to select objects of interest, which can then be used to aid image editing.
Both the user study and the large scale quantitative evaluation verify the usefulness of
our verbal parsing method.
Our verbal interaction is especially suitable for new generation devices
such as smart phones, Google Glass, consoles and living room devices.
To encourage the research in this direction,
we will release source code and benchmark datasets.

\myPara{Limitations} Our approach has some limitations.
Firstly, our reliance on attribute handles can fail if there is
no combination of attributes that can be used to improve the image parsing.
This can seen in the second and eighth image of \figref{parsed} where we fail to
provide any verbal refined result due to lack of appropriate attributes.
Of the $78$ images we tested ($55$ from dataset and $23$ Internet images)
only $10$ ($5$ dataset and $5$ Internet images) could not be further refined using attributes.
This represents a $13\%$ failure rate.
Note that refinement failure does not imply overall failure and
the automatic results may still be quite reasonable as seen in \figref{parsed}.
Secondly, the ambiguity of language description
prevents our algorithm from giving $100\%$ accuracy.

\myPara{Future work}
Possible future directions might include extending our method
to video analysis and inclusion of stronger physics based models
as well as the use of more sophisticated techniques from machine learning.
Interestingly our system can often segment objects that are not
in our initial training set by relying solely on their attribute descriptions.
In the future, we would like to better understand this effect and
suitably select a canonical set of attributes to strengthen this functionality.
It might also be interesting to explore efficient multi-class object
detection algorithms to help working set selection,
possibly supporting thousands of object classes \cite{google10K_Class,BingObj2014}.
We have only scratched the surface of verbal guided image parsing
with many future possibilities,
e.g., how to better combine touch and verbal commands, or
how verbal refinement may change the learned models
so that they perform better on further refinements.

\section*{Acknowledgements}
We would like to thank the anonymous associate editor and reviewers for their
valuable feedbacks,
and Michael Sapienza for the voice over in the video.
This research was supported by EPSRC (EP/I001107/1), ERC-2012-AdG 321162-HELIOS,
and ERC Starting Grant SmartGeometry 335373.

\bibliographystyle{acmsiggraph}
\bibliography{ImageSpirit}
\end{document}